\documentstyle[12pt]{article}
\begin{document}
\newcommand{\scr}{\sin^2 \hat{\theta}_W (m_Z)}
\newcommand{\sef}{\sin^2 \theta_{eff}^{lep}}
\newcommand{\msbar}{\rm{\overline{MS}}}
\newcommand{\be}{\begin{eqnarray}}
\newcommand{\en}{\end{eqnarray}}
\newcommand{\kcar}{\hat{k}}
\newcommand{\scar}{\hat{s}}
\newcommand{\cc}{\hat{c}}
\newcommand{\alc}{\hat{\alpha}}
\newcommand{\nn}{\noindent}

\nn     \hspace*{10.6cm} NYU--TH--93/09/07 \\
\hspace*{11.3cm} September 1993 \\

\vspace*{3.5cm}

\centerline{\large{\bf Relation between $\scr$ and $\sef$. }}

\vspace*{1.5cm}

\centerline{\sc Paolo Gambino and    Alberto Sirlin.}

\vspace*{1cm}

\centerline{ Department of Physics, New York University, 4 Washington
Place,}
\centerline{ New York, NY 10003, USA.}

\vspace*{3cm}

\begin{center}
\parbox{14cm}
{\begin{center} ABSTRACT \end{center}
\vspace*{0.2cm}
The relation between $\sef$, frequently employed in  LEP analyses,
and the
$\rm{\overline{MS}}$--parameter  $\scr$ is discussed and their
difference evaluated by means of an explicit calculation.  }

\end{center}
\vfill
\newpage
\nn It has been emphasized that the $\msbar$ parameter $\scr$ provides
a very convenient framework to discuss physics at the $Z^0$ peak
\cite{1} and, as it is well known, it plays a crucial r$\rm{\hat{o}}$le
in the analysis of grand unification. On the other hand, the
LEP collaborations employ an effective coupling, $\sef$ \cite{2,lep}.
 It is a
 common belief among physicists that these two parameters, although
very different conceptually, are very close numerically.
However, the reason and extent of this coincidence and the precise
conceptual
and numerical relation between the two  has not been spelled out in the
literature. In turn this is a source of considerable confusion among
theorists and experimentalists alike. The aim of this report is to
clarify these issues.

 The effective weak interaction angle employed by the LEP groups is
defined by
\be
1-4\sef=\frac{g_V^{\ell}}{g_A^{\ell}}
\label{def}
\en
where $g_V^{\ell}$ and $g_A^{\ell}$ are the effective vector and
axial couplings in the $Z^0\rightarrow\ell\bar{\ell}$
amplitude at resonance, where $\ell$ denotes a charged lepton
\cite{2,lep}. In order to establish the connection with $\scr$,
we note that this amplitude is proportional to \cite{4}
\be
<\ell\bar{\ell}|J_Z^\lambda|0>= -\bar{u}_\ell\gamma^\lambda
\left[\frac{1-\gamma_5}{4} - \kcar_{\ell}(q^2)\ \scar^2\right]v_{\ell},
\label{ampl}\en
where $ \scar^2$ is an abbreviation for $\scr$, $v_{\ell}$
and $\bar{u}_{\ell}$ are the lepton spinors and $\kcar_{\ell}(q^2)$
is an electroweak form factor.
Up to terms of order $\cal{O}(\alpha)$ we have \cite{4}
\be
\kcar_\ell(q^2)=1-{\cc\over\scar}\frac{\left[
A_{\gamma Z}(q^2)-A_{\gamma Z}(0)\right]_{\msbar}}{q^2-A^{(f)}_{
\gamma\gamma\msbar}(q^2)}
+\frac{\alc}{\pi \scar^2}\cc^2\log c^2
-{\alc\over4\pi \scar^2}V_\ell(q^2),
\label{kappa}
\en
where $A_{\gamma Z}(q^2)$ is the  $\gamma$--$Z$ mixing self--energy,
the subscript $\msbar$ means that the $\msbar$ renormalization has been
carried out (i.e. the pole terms have been subtracted and the 't--Hooft
scale has been set equal to $m_Z$),
the superscript $f$ stands for fermionic part,
$\alc$ is an abbreviation for
$\alc(m_Z)=[127.9\pm 0.1]^{-1}$ \cite{6},
$\cc^2\equiv \cos^2\hat{\theta}_W(m_Z)$,
$c^2\equiv m_W^2/m_Z^2$ and $V_\ell(q^2)$ is a finite vertex
correction. Explicitly,
\be
V_\ell(q^2)={1\over2}f({q^2\over m_W^2}) +4\cc^2\  g({q^2\over m_W^2})-
\frac{1-6\scar^2 +8\scar^4}{4\cc^2} f({q^2\over m_Z^2}),
\label{vertex}
\en
where $f(x) $ and $ g(x) $ are defined in Eqs. (6d, 6e) of Ref.\cite{4}.
We have included the photon self--energy $A^{(f)}_{\gamma\gamma\msbar}
(q^2)$
in the second term of Eq.(\ref{kappa}) because, as it will be explained
later,
it gives rise to relatively large ${\cal O}(\alpha^2)$ terms.

It is clear from Eq.(\ref{ampl}) that the ratio of the vector and
axial vector couplings at resonance is given by
$1-4\ \kcar_\ell(m_Z^2)\ \scar^2$.
We now discuss the various contributions to $\kcar_\ell (m_Z^2)$.

To ${\cal O}(\alc)$ the fermionic contribution
to the real part of Eq. (\ref{kappa}) can be
written in the form
\be
-\frac{\cc}{\scar}\frac{{\rm{Re}} \ A_{\gamma Z}^{(f)}(m_Z^2)}{m_Z^2}
={\hat{e}^2\over\scar^2}\sum_i \left(\frac{Q_i C_i}{4} -\scar^2
Q_i^2\right)
\frac{{\rm{Re}}\ \Pi^V(m_Z^2, m_i,m_i)}{m_Z^2},
\label{agz}
\en
where $Q_i$, $C_i$, and $m_i$ are the charge, third component of
weak isospin (with eigenvalues $\pm1$), and  mass  of the $i$--th
fermion, the summation includes the color degree of freedom, $\Pi^V$ is
the vacuum polarization function involving vector currents, and
henceforth the $\msbar$ renormalization is not indicated explicitly.
 For the leptons we set $m_i=0$ and find that the contribution to
 Eq.(\ref{agz})
is $(\alc/\pi \scar^2)(5/12)(1-4\scar^2)=3.2\times 10^{-4}\ $\cite{5}.
In this calculation and henceforth we employ $\scar^2=0.2323$,
which corresponds to the central values $m_t=162$ GeV and $m_H=300$
GeV in the global fit of Ref.\cite{lep}, and $\alc=(127.9)^{-1}$.

For the first five quark flavors we again set $m_i=0$ and, including
${\cal{O}}(\alc \alc_s)$ corrections, obtain a contribution
\cite{6} $(\alc/\pi\scar^2)\ (7/12 - 11\scar^2/9)\ [5/3 +
(\alc_s/\pi)(55/12-4\zeta(3))]=5.32\times 10^{-3}$,
where we have used $\alc_s=\alc_s(m_Z)=0.118$ and $\zeta(3)=1.20206...$.

The top quark contribution to Eq.(\ref{agz}) is of the form \cite{6}
\be
-{\cc\over\scar}
\frac{{\rm{Re}}\ A_{\gamma Z}^{(top)}(m_Z^2)}{m_Z^2}=
-{\alc\over 6\pi\scar^2}
\left(1-{8\over3}\scar^2\right)\left[\left(
1+{\alc_s\over\pi}\right)\log\xi_t - {15\over4}{\alc_s\over\pi}\right]
+D({1\over\xi_t}),
\label{top}\nonumber\\
\en
where $\xi_t\equiv m_t^2/m_Z^2$ and $D(1/\xi_t)$ represents small
terms that decouple in the limit $\xi_t\rightarrow\infty$.
For the current range $120$ GeV ${<\atop\sim}
m_t {<\atop\sim} 200$ GeV \cite{lep},
$D(1/\xi_t)$ varies from $1.0\times 10^{-4}$ to $3\times 10^{-5}$
and is of the
 same order of magnitude as neglected two--loop contributions
$\sim (\alc/\pi \scar^2)^2\approx 10^{-4}$ to Eq.(\ref{kappa}).

According to the Marciano--Rosner [M--R]
convention \cite{mr}, adopted also in Ref.\cite{6},
the first term in Eq.(\ref{top}) is subtracted  in the evaluation
of ${\rm Re}\ A_{\gamma Z}^{(top)} (m_Z^2)/$ $m_Z^2$.
This is part of the
$\msbar$ renormalization prescription of these authors,
 the idea being that contributions from particles
of mass $m>m_Z$ that do not decouple in the limit $m\rightarrow
\infty$ are subtracted from this particular amplitude and absorbed
 in the definition of $\scr$.
The aim of the prescription is to make the value of $\scr$, as
extracted from
the on--resonance asymmetries, very insensitive to heavy particles of
mass $m>m_Z$. We reach the conclusion that when the M--R
prescription is
applied, the top quark contribution to Eq.(\ref{agz}) is
very small.

The other contributions to $ \kcar_\ell(m_Z^2)$
in Eq.(\ref{kappa})
can be readily obtained from the literature.
This form factor is gauge invariant, but several individual components
are not. We evaluate them in the 't Hooft--Feynman gauge, using
$m_W=80.23$ GeV \cite{lep}:
 ($i$) the bosonic contributions
$-(\cc/\scar)\ [A_{\gamma Z}^{(b)} (m_Z^2)-A_{\gamma Z}^{(b)}(0)]/m_Z^2$
can be extracted from Ref.\cite{consid} and amount to
$-5.92\times 10^{-3}$;
($ii$) $-(\alc/4\pi\scar^2)\ {\rm Re}\ V_\ell(m_Z^2)$ can be obtained
 from
Eq.(\ref{vertex}) of this paper and Eqs.(6d,e) of Ref.\cite{4},
and gives $+3.32\times 10^{-3}$;
($iii$) $(\alc/\pi \scar^2)\cc^2\log c^2= -2.11\times 10^{-3}$;
($iv)$ although two--loop effects have not been fully
calculated, we  include the ${\cal O}(\alc^2)$ contribution arising
 from the product of ${\rm Im}A_{\gamma Z}(m_Z^2)$ and
${\rm Im}A_{\gamma\gamma}(m_Z^2)$ in the second  term of
Eq.(\ref{kappa}). It amounts to $+1.9\times
10^{-4}$. It is quite sizeable, relative to typical ${\cal O}(\alc^2)$
contributions, because these imaginary parts involve  several
additive terms. On the other hand, the large logarithmic
${\cal O}(\alpha^2)$ corrections associated with the running of
$\alpha$
are already taken into account, in the $\msbar$ scheme, by employing
$\alc$ in the evaluation of $A_{\gamma Z}$;
$(v)$ for  $120$ GeV ${<\atop\sim} m_t {<\atop\sim} 200$ GeV,
the $t - \bar t$
threshold contribution \cite{6,9} to Eq.(\ref{agz}) ranges from $1.7
\times 10^{-5} $ to $ 2.7
\times 10^{-5}$ and is therefore negligible;
 ($vi$) there are imaginary contributions to $\kcar_\ell(m_Z^2)$
arising from $A_{\gamma Z}^{(f)}(m_Z^2)$ and $V_\ell (m_Z^2)$ and
amount to $i\ 1.06\times 10^{-2}$ and  $i\ 0.28\times 10^{-2}$,
respectively.

Combining all the above results we have
$ \kcar_\ell(m_Z^2)= 1\ + \ (0.32+5.32-5.92+3.32-2.11+0.19)
\times 10^{-3}
+D(1/\xi_t)+ i\ (1.06+0.28)\times 10^{-2}$,
which to good approximation becomes
\be
\kcar_\ell(m_Z^2)=1.0012\ +\ i\ 0.0134.
\label{kcar}\en

It is clear, on the basis of Eq.(\ref{kcar}), that at the
one--loop level
the ratio of  effective vector and axial vector couplings in the
 $Z\rightarrow \ell \bar\ell$ amplitude is a complex number. This
 is also expected from general principles. On the other hand, the LEP
groups interpret both sides of Eq.(\ref{def}) as real quantities.
This can be justified on the grounds that the imaginary component of
$\kcar_\ell (m_Z^2) $ gives a negligible contribution to the leptonic
bare asymmetries and partial widths.
For instance, the bare left--right asymmetry is given by
$A^{0,\ell}_{LR}= \ 2 {\rm Re}(g_V/g_A)/[1\ +\ |g_V/g_A|^2]$   and one
readily verifies that
 the inclusion of ${\rm Im}\ \kcar_\ell
(m_Z^2)$ decreases its value
by only $-0.02\%$. Similarly, $A^{0,\ell}_{FB}$ is modified by
$\approx -0.03\%$.
Therefore we identify
\be
\sef=\scar^2 \ {\rm Re}\ \kcar_\ell(m_Z^2).
\label{ }
\en
Using Eq.(\ref{kcar}) we have
\be
\sef-\scr=2.8\times 10^{-4}\approx 3\times 10^{-4}.
\label{res}
\en

The following observations are appropriate at this stage:
$(a)$ because the Higgs boson does not contribute
 at the one--loop level
to Eq.(\ref{kappa}), the results of Eqs.(\ref{kcar}, \ref{res}) are
 independent of $m_H$;
$(b)$ it is clear that the closeness of ${\rm Re}\ \kcar_\ell(m_Z^2)$
to unity and, correspondingly, the small difference in
Eq.(\ref{res}) are due
to the cancellation of significantly larger terms. For instance,
the light quark and bosonic contributions to the $\gamma$--$Z$
mixing self--energy are of the roughly expected order of magnitude
$\sim \alc/(2\pi\scar^2)\approx 5.4\times 10^{-3}$, but they
 largely cancel
against each other. On the other hand, the ${\cal O} (\alc)$
contributions to the ${\rm Im}\ \kcar_\ell (m_Z^2)$
are $\approx 1\%$, an order of magnitude larger;
(c) as the relation between $\scr$ and $\sin^2 \theta_W\equiv
1-m_W^2/m_Z^2$
is well--known \cite{6}, Eq.(\ref{res}) determines the connection
between the three parameters.

If the Marciano--Rosner decoupling convention is not applied,
so that in the
$\msbar$ renormalization  one only subtracts poles and sets the
 't--Hooft
scale equal to $m_Z$, there is a further contribution  to ${\rm Re}\
\kcar_\ell(m_Z^2)$ arising from the first term in Eq.(\ref{top}).
Using $\alc_s(m_t)\approx 0.11$, this amounts to $-3\times 10^{-4}$,
 $-7\times 10^{-4}$, $-1.0\times 10^{-3}$,
for $m_t= 120,\ 162$ and 200 GeV, respectively. Correspondingly,
${\rm Re}\ \kcar_\ell(m_Z^2)$ becomes 1.0009, 1.0005, 1.0002, even
closer to unity.
As a consequence, although the difference between $\sef$ and $\scr$
depends more on $m_t$ when the M--R prescription is not applied,
it is actually smaller for the current range $120 GeV
{<\atop\sim} m_t {<\atop\sim} 200 GeV$.
In fact, we find that it is $+2\times 10^{-4}$ for $120\le
 m_t\le 135$ GeV,
$1\times 10^{-4}$ for $136\le mt \le 184$ GeV, and there is
no difference in
the fourth decimal for  $185\le m_t\le 200$ GeV.

One can obtain a rough consistency check of the order of magnitude of
Eq.(\ref{res}) by comparing the fits of Ref.\cite{lep} with
the calculations of
Ref.\cite{6}. Using the LEP,  collider and $\nu$ data, Ref.\cite{lep}
finds $m_t=162{+16\atop -17}{+18\atop -21}$ and $\sef=.2325\pm 0.0005
{+0.0001\atop -0.0002}$ for constrained $\alc_s$,
and the same value of $m_t$ but a slightly different central value
for $\sef$ (0.2326), with the same errors, in their unconstrained
$\alc_s$ fit.
 Their central values assume $m_H=300$ GeV, the first
error represents experimental and theoretical uncertainties,
while the second
reflects changes corresponding to the assumptions $m_H=60$ GeV
 and $M_H=1$ TeV.
According to Eq.(\ref{res}), the corresponding central values
for $\scar^2$ should be 0.2322 and 0.2323.
On the other hand, from Ref.\cite{6} one finds $\scar^2=0.2323$
for $m_t=162$ GeV and $m_H=300$ GeV. Thus, the comparison of
the conclusions of Ref.\cite{lep} with the calculations of
Ref.\cite{6} is roughly consistent with
Eq.(\ref{res}).
Of course, such rough consistency checks are not a substitute for
precise,
ab initio calculations, like the one leading to Eq.({\ref{res}}).

In order to avoid possible further sources of confusion, we make two
additional comments.
 ($a$) Sometimes, consistently with Eq.(\ref{def}), rapporteurs define
$\sef$ in terms of a bare forward--backward asymmetry $A^{0,\ell}_{FB}$,
which is obtained from the physical asymmetry $A^\ell_{FB}$ after
 extracting the effect of photon--mediated contributions
and other radiative correction effects \cite{2}.
 Therefore, we should not attempt to find the numerical relation
between $\sef$ and $\scr$ by comparing detailed $\msbar$ calculations
of the physical asymmetry $A_{FB}^\ell$,
as those in Ref. {4 }, with theoretical
expressions for $A^{0,\ell}_{FB}$ expressed in terms of $\sef$.
The point is that $A_{FB}^\ell$ contains electroweak effects not
contained in $A_{FB}^{0,\ell}$.
($b$) Rapporteurs often cite the value of $\sef$ as extracted only from
 the on--resonance asymmetries, while they give the prediction for $m_t$
derived from the complete data base.
Current asymmetry results lead to determinations of $\sef$ that are
somewhat smaller than the $\scr$ numbers corresponding to
the central $m_t$. This, however, is not a contradiction
with Eq.(\ref{res}), because the on--resonance asymmetries represent
 only
 a part of the experimental information.
 This is quite visible in the detailed report of  Ref.
\cite{lep}, in which one finds  $\sef=0.2321\pm 0.0006$ from
the on--resonance asymmetries and, as mentioned before, larger
values from the global fits.

In summary, we have attempted to clarify the connection
between $\ \sef$ and $\ \scr$
and obtained the value of their difference by
means of a detailed calculation, both with and without the M--R
decoupling convention. In view of the prospects for a very
accurate determination of the mixing angle from further
LEP studies and from $A_{LR}$ at SLAC,
 and the fact that both definitions are frequently
employed, we feel that this clarification is indeed timely.
\vskip 1.5cm

\noindent{\bf{Aknowledgements}}
\vskip .4cm
One of us (A. S.) would like to thank Prof. W. Hollik for  very useful
conversations on this subject.
This research was supported in part by the NSF under Grant No.
PHY--9017585.

\end{document}